# Meeting experimental challenges to physics of network glasses: assessing role of sample homogeneity

*S. Bhosle, K. Gunasekera, P. Chen, P. Boolchand*
Department of Electrical and Computer Engineering
University of Cincinnati, Cincinnati, OH 45221-0030

*M.Micoulaut*
Laboratoire de Physique Theorique de la Matiere Condensee, Universite Pierre et Marie Curie
Boite 121, 4 Place Jussieu, 75252 ParisC Cedex 05, France

*C. Massobrio*
Institut de Physique et de Chimie des Materiaux de Strasbourg, 23 rue du Loess, BP43, F-67034
Strasbourg Cedex 2, France

We introduce a Raman profiling method to track homogenization of $Ge_xSe_{100-x}$ melts in real time, and show that 2 gram melts reacted at 950°C in high vacuum homogenize in 168 hours on a scale of 10μm. Homogenization of melts is precursive to self-organization of glasses. In the present glasses, compositional variation of Raman active corner-sharing mode frequency of $GeSe_4$ units, molar volumes, and the enthalpy of relaxation at $T_g$ , reveal the rigidity ( $x_c(1)$= 19.5(3)%) and the stress ($x_c(2)$ = 26.0(3)%) transitions to be rather sharp ($\Delta x < 0.6\%$). These abrupt elastic phase transitions are *intrinsic* to these materials and have a direct bearing on physics of glasses.

Bulk glasses have been synthesized by reacting starting materials to produce melts, which are then quenched to produce disordered solids[1]. One generally assumes that holding melts at several hundred degrees above the liquidus for several hours would homogenize them. In equilibrium phase diagrams, glass forming compositions are usually bordered by congruently melting crystalline phases[2]. These crystalline phases, in principle, can nucleate as melts are quenched to produce microscopic heterogeneities (MH). In practice, such equilibrium thermodynamic effects can be suppressed in *pure* melts as heterogeneous nucleation sites become minuscule as shown recently in metallic glasses[3]. We introduce here a Raman profiling method to track homogenization of chalcogenides in real time, and find that the process consists of two steps, an initial *step 1* as starting materials react to produce MH consisting of crystalline phases embedded in a glass. Continued reaction leads the crystalline phases to dissolve and local structures characteristic of melts/glasses to emerge. In *step 2,* intermediate and extended range structures evolve as melt stoichiometry across a batch composition equalizes by a process described as Melt-Nanoscale Mixing (MNM). We illustrate these ideas for the case of the well studied $Ge_xSe_{100-x}$ binary[4-14], and find that *slow* MNM and not MH due to thermodynamic phase separation[15] is the determinative factor that contributes to glass heterogeneity in chalcogenides.

The finding has a direct bearing on the sharpness of the *rigidity* and *stress* transitions in the $Ge_xSe_{100-x}$ binary system. Rigidity theory [16-18] has been the corner stone to understanding network glasses in terms of their topology. An intermediate phase (IP) forms between these transitions [19]and has attracted widespread interest because of its exceptional functionalities, including the stress-free, non-aging, dynamically reversible nature of networks formed in this nanostructured phase[8, 9, 18, 20, 21]. In the present homogenized glasses, we find the rigidity and stress transitions to occur near x = 19.5(3)% (*rigidity*) and 26.0(3)% (*stress*) and to be remarkably sharp (width $\Delta x < 0.6\%$) . Our findings illustrate that the intrinsic stress-driven behavior of these chalcogenide glasses may be far richer than hitherto recognized. The Raman profiling method provides a powerful means to access homogeneous glasses and melts to explore their intrinsic nanostructure.

Elemental Ge and Se lumps (3-4 mm diam.) of 99.999% purity from Cerac Inc, encapsulated in 5mm ID quartz tubing at $1x10^{-7}$ Torr, were reacted at 950°C for periods, $t_R$, ranging from 6h < $t_R$ < 168 h in a box furnace, with tubes held vertically. Batch sizes were kept near 2 grams. Prior to use, quartz tubing was dried in a vacuum oven (90°C) for 24 hours. Periodically, melt temperatures were lowered to 50°C above liquidus[2] and water quenched, and examined in FT-Raman profiling experiments. These measurements used 1.064 µm radiation from a Nd-YAG laser with a 50µm spot size to excite the scattering, and spectra were acquired along the 1 inch length of a glass column at 9 locations ( Fig.1). In the initial stages of alloying ($t_R$ = 6 hours), melts are, indeed, quite heterogeneous as revealed by significant changes in the Raman spectra from point 1 to 9. The 9 Raman lineshapes are superimposed in Fig 2a, and provide a pictorial view of glass heterogeneity. Continued reaction ($t_R$ = 96h) of melts, increases homogeneity (Fig 2c), but a fully homogeneous melt (glass) is realized only after $t_R$ = 168h (Fig 2d) when all 9 line shapes coalesce. We have synthesized 21 glass compositions in the 10% < x < 33.33% range, and ascertained their homogeneity by Raman profiling scans in each case . Separately, we also synthesized glasses at x = 19% and 33.33% using finely crushed Ge and Se powders stored at laboratory ambient (45% rel. humidity) as starting materials. These melts reacted quicker in the first step ( $t_R$< 48h) but took as much time ($t_R$ = 96 hours) as others to nanoscale mix in step 2. However, their structural properties are measurably different from their dry counterparts and are characteristic of wet samples containing hydrolyzed products (see below). The final step in synthesis was to thermally cycle all samples through $T_g$ and slow cool to room temperature at 3°C/min to remove stress frozen upon a water quench.

In step 1 of reaction for the case of a melt at x = 19% , MH are first manifested as α-$GeSe_2$ [6] fragments nucleate in the glass at the tube bottom, and the evidence consists of the narrow modes (arrows) observed at locations 1,2, and 3 in Figs. 1, and Fig 2a and b. At point 4 in Fig 1, a binary glass of $Ge_{19}Se_{81}$ stoichiometry forms, and as one approaches the top (point 9) of the column, melts become steadily Ge-deficient ($Ge_8Se_{92}$), as estimated from the increased scattering strength ratio of the chain-mode (CM, near 250 $cm^{-1}$) to corner- sharing mode (CS, near 200 $cm^{-1}$) [6, 7]. With increasing $t_R$ > 24h the crystalline phase dissolves. At $t_R$ = 96h, the crystalline phase vanishes, but a heterogeneous melt persists (Fig 2c) with Ge content 'x' varying almost linearly from 21% at location 1 to 17% at location 9 along the length of the column. In the spectra, the absence in the spatial variation of the CS mode strength is due to normalizing the spectra to that mode. At this point appropriate local structures of melts /glasses have evolved, and further reaction of the melt to $t_R$ = 168 h leads to the Ge content across the batch composition equalizing, and a fully homogeneous glass to be realized on a scale of 10μm or less (deduced from micro-Raman experiments). The 2-step behavior of homogenization of melts reported here at x = 19% is observed at all other compositions examined in the present $Ge_xSe_{100-x}$ binary. Our experiments also reveal that rocking the reaction tube speeds up step 1 of the homogenization, but it is step 2 of MNM that is the rate limiting process to melt homogenization. MNM requires a large number of correlated sequential bond-breaking and bond-forming steps for the 4-fold coordinated Ge to diffuse, and the Ge/Se ratio across a batch to equalize. Separately, 1/4 gram sized melts were also studied, and found to homogenize in 6 h rather than 168 h needed for the 2 gram batch size. The result is the consequence of a 5-fold reduction in diffusion length for Ge and Se atoms to move across as concentration gradients vanish. From these data, we obtain a Diffusion constant for Ge and Se atoms in $Ge_xSe_{100-x}$ melts at 950°C of $D = 4 \times 10^{-6} cm^2/sec$.

Dispersive Raman system (Model T64000 system from Horiba Inc.) using 5 mW of 647 nm radiation with a 50 μm laser spot size was used to excite the scattering of homogenized glasses encapsulated in quartz tubes. The observed lineshapes were least-squares fit [7] to superposition of Gaussians, and the variation in the CS mode frequency ( $\nu_{CS}(x)$) deduced (Fig 3a). These data show three distinct regimes of variation, a power-law behavior at x > 26.0%, an approximately linear variation in the 19.5% < x < 26% range, and again a linear variation but with a higher slope at x < 19%. Density of glasses could be measured using a quartz fiber with a digital balance and dry alcohol to 1/4% accuracy with large samples. The variation of molar volumes ($V_m(x)$) with composition (Fig 3b) on dry glasses also show three distinct regimes- a nearly flat regime in the 19.5% < x < 26% range, and a rapid rise at x > 26% and at x < 19% , with the two thresholds coinciding with the Raman $\nu_{CS}(x)$ trends. Glass transition temperatures, $T_g(x)$, the jump in the specific heat at $T_g$ ($\Delta C_p(x)$ ) and the enthalpy of relaxation at $T_g$ ( $\Delta H_{nr}$ (x)) [7], were also measured using a model 2920 mDSC from TA Instruments, and some of these results appear in Fig 3a and c.

The observed CS mode frequency variation, $\nu_{CS}(x)$ (Fig.3a) , serves to uniquely identify the phase formed at x > 26% to be the *stressed-rigid* phase [7]. We have extracted the underlying optical elasticity (which varies as $\nu_{CS}^2$) power-law variation in x, using equation (1)

$$\nu_{CS}^2(x) - \nu_{CS}^2(x_c(2)) = A ( x - x_c(2))^{p_2} \qquad (1)$$

Here $\nu_{CS}^2(x_c(2))$ represents the value of $\nu_{CS}^2(x)$ at the threshold composition, $x = x_c(2)$, and $p_2$ the elastic power law in the *stressed-rigid* regime. The data at x > 26% (Fig.3a) was used to extract $p_2$, this time by an *iterative process* using both a polynomial fit and separately a log-log fit to (1). The value of $x_c(2)$ was varied so that both fitting procedures yielded the same $p_2$, and the final result (Fig 4a) gives $p_2 = 1.50(3)$ and $x_c(2) = 26.0(3)$. The value of $p_2$ is in excellent agreement with a numerical simulation of the power-law[22]. For the IP, a similar procedure gives (Fig 4b) $x_c(1) = 19.5(3)\%$ and $p_1 = 1.10$ (5). The present value of $p_1$ is larger and more accurate than our previous report[19] and its magnitude almost identical to values noted earlier in IPs of modified oxides[23] and chalcogenides[24].

Most striking is the sharpening of the *rigidity* and *stress* transitions upon *aging* of samples. The $\Delta H_{nr}(x)$ results on fresh samples (curve F, Fig 3c) become *step- like* near x = 19.5% after 2 weeks of aging at room temperature ($T_{aging}$ = 23°C) (curve A1,Fig.3c). Use of a higher $T_{aging}$ = 240°C for stressed-rigid glasses ( because of their higher $T_g$) also leads to a striking *step-like* increase of $\Delta H_{nr}(x)$ term near the stress-transition (curve A2, Fig 3c). These A1 and A2 data sets on *aged* samples must be compared with the triangular variation of $\Delta H_{nr}(x)$ observed on 2- week aged samples (Curve A0,Fig 3c) reported in ref. [7]. We can now estimate the spread in Ge stoichiometry of glasses in ref [7] to be + or – 2% in x from the known $t_R$. For the composition close to the reversibility window center, x ~ 23%, one expects both data sets ( A0 and A1) to show the $\Delta H_{nr}(x)$ term to vanish, as they indeed do. However, as one goes away from the center, the $\Delta H_{nr}(x)$ term should increase linearly in the heterogeneous samples (of ref 7) as contributions to the heat flow term from the flexible (stressed-rigid) phase steadily weigh in on the low (high) x side. Thus, one can naturally understand how square-well like variation of $\Delta H_{nr}(x)$ in the present very homogeneous samples translates into an almost triangular (fig 3c) variation in the heterogeneous ones of ref.[7]. Flexible and stressed-rigid structures compact upon aging and lower the entropy of a glass as found at x < 20% and at x > 26%, and lead the $\Delta H_{nr}$ term to increase. The sharpening of the *rigidity* and *stress* transitions upon aging is a natural consequence of compositions outside the IP aging but those in the IP barely age.

Of more than passing interest is the lack of variation in $\Delta C_p(x)$ term in the present homogeneous glasses as a function of x.(Fig 3c) Such a behavior was also noted in the $Ge_xAs_xSe_{100-2x}$ ternary [25]. In the present binary [26] and the GexAsxSe100-2x ternary [27], the fragility of melts shows a minimum for IP compositions. Thus, it appears that for the case of chalcogenides that the fragile-strong classification of melts ($T > T_g$) [28] based on dynamics (viscosity) correlates much better with the variation in $\Delta H_{nr}(x)$ term than with the $\Delta C_p(x)$ term (Fig 3c) in glasses ( $T < T_g$).

The molar volume trends, $V_m(x)$ (Fig 3b) suggest that glass samples of ref [5] are not as dry as those of ref [4], while those of ref [4] not as homogeneous at the present ones. These observations highlight the need to seal the pure and dry starting materials as lumps in quartz tubes under high vacuum (< $10^{-7}$ Torr) to avoid bonded water related artifacts.

The nature of the sharp threshold observed near the composition $x_c(3)$ = 31.5% (Fig.3) deserves a final comment. In Raman scattering, Ge-Ge bonds as part of ethane-like units[10] first manifest near x = $x_c(3)$ = 31.5%. The cusp in $\Delta H_{nr}(x)$ (Fig3c) coincides with a maximum in the slope $dT_g/dx$ (Fig 3a). Both these observables are related to the network topology.[29]. And we understand the reduction in $\Delta H_{nr}(x)$ and in $dT_g/dx$ at x > 31.5% as due to the decoupling[10] of ethanelike units from the backbone. The nanoscale phase separation leads to a maximum[10] of $T_g$ near x = 33.33%.

In summary, a Raman profiling method has permitted synthesis of bulk $Ge_xSe_{100-x}$ glasses of unprecedented homogeneity, resulting in sharply defined rigidity and stress transitions. These considerations will apply generally to other chalcogenides. Melt homogenization on a scale of 10 μm appears sufficient to promote self-organization of chalcogenide glasses, and opens a new avenue to experimentally access the *intrinsic* physical behavior of these fascinating materials, both in the glassy ($T < T_g$) and the liquid [30, 31] state ($T > T_g$). We thank D. McDaniel, L. Thomas , B .Goodman, and B. Zuk for discussions. This work is supported by NSF grant DMR 08-53957.

## References

[1] Zallen R 1983 *The Physics of Amorphous Solids* (New York: Wiley).

[2] Ipser H, Gambino M and Schuster W 1982 *Monatshefte für Chemie / Chemical Monthly* **113** 389.

[3] Yamamoto T, Yodoshi N, Bitoh T, Makino A and Inoue A 2008 *Reviews on Advanced Materials Science* **18** 126.

[4] Mahadevan S, Giridhar A and Singh A K 1983 *Journal of Non-Crystalline Solids* **57** 423.

[5] Feltz A, Aust H and Blayer 1983 *Journal of Non-Crystalline Solids* **55** 179.

[6] Wang Y, Nakamura M, Matsuda O and Murase K 2000 *Journal of Non-Crystalline Solids* **266-269** 872.

[7] Feng X W, Bresser W J and Boolchand P 1997 *Physical Review Letters* **78** 4422.

[8] Chen G, Inam F and Drabold D A 2010 *Applied Physics Letters* **97** 131901.

[9] Massobrio C, Celino M, salmon P S, Martin R A, Micoulaut M and Pasquarello A 2009 *Physical Review B* **79** 174201.

[10] Boolchand P and Bresser W J 2000 *Philosophical Magazine B* **80** 1757.

[11] Gjersing E L, Sen S and Aitken B G 2010 *The Journal of Physical Chemistry C* **114** 8601.

[12] Inam F, Chen G, Tafen D N and Drabold D A 2009 *Physica Status Solidi B - Basic Solid State Physics* **246** 1849.

[13] Salmon P S 2007 *Journal of Non-Crystalline Solids* **353** 2959.

[14] Petri I, Salmon P S and Fischer H E 2000 *Physical Review Letters* **84** 2413.

[15] Lucas P, King E A, Gulbiten O, Yarger J L, Soignard E and Bureau B 2009 *Physical Review B* **80** 214114.

[16] Phillips J C 1979 *Journal of Non-Crystalline Solids* **34** 153.

[17] Thorpe M F, Jacobs D J, Chubynsky M V and Phillips J C 2000 *Journal of Non-Crystalline Solids* **266** 859.

[18] Micoulaut M and Phillips J C 2003 *Physical Review B* **67** 104204.

[19] Boolchand P, Feng X and Bresser W J 2001 *Journal of Non-Crystalline Solids* **293** 348.

[20] Brière M A, Chubynsky M V and Mousseau N 2007 *Physical Review E* **75** 056108.

[21] Barre J, Bishop A R, Lookman T and Saxena A 2005 *Physical Review Letters* **94** 208701.

[22] He H and Thorpe M F 1985 *Physical Review Letters* **54** 2107.

[23] Rompicharla K, Novita D I, Chen P, Boolchand P, Micoulaut M and Huff W 2008 *Journal of Physics: Condensed Matter* **20** 202101.

[24] Qu T, Georgiev D G, Boolchand P and Micoulaut M 2003 The intermediate phase in ternary $Ge_xAs_xSe_{1-2x}$ glasses *Supercooled Liquids, Glass Transition and Bulk Metallic Glasses* vol 754 (Warrendale, PA: Material Research Society) pp. 157.

[25] Wang Y, Boolchand P and Micoulaut M 2000 *Europhysics Letters* **52** 633.

[26] Stolen S, Grande T and Johnsen H-B 2002 *Physical Chemistry Chemical Physics* **4** 3396.
[27] Böhmer R and Angell C A 1992 *Physical Review B* **45** 10091.
[28] Angell C A 2000 Glass formation and the nature of the glass transition *Insulating and Semiconducting Glasses* ed P. Boolchand (SinRiver Edge, NJ: World Scientific) pp. 1-51.
[29] Micoulaut M and Naumis G G 1999 *Europhysics Letters* **47** 568.
[30] Bermejo F J, Cabrillo C, Bychkov E, Fouquet P, Ehlers G, auml, ussler W, Price D L and Saboungi M L 2008 *Physical Review Letters* **100** 245902.
[31] Mauro J C and Loucks R J 2008 *Physical Review E* **78** 021502.

## Figure Captions

**Fig.1.** Raman scattering of quenched $Ge_{19}Se_{81}$melt taken along the quartz tube length at 9 locations, 6 hours after reacting the starting materials at 950°C. The sharp modes at the arrow locations are those of 2D or α-$GeSe_2$. CS = Corner Sharing, ES = Edge Sharing and CM = Chain mode.

**Fig.2.** A coalesced view of the 9 Raman spectra of Fig.1 appears in panel (a). Prolonged reaction of the $Ge_{19}Se_{81}$melt for (b) 24h, (c) 96h , (d) 168h (d) show it homogenizing. In (c), we provide the color versus location key of in Fig1.

**Fig.3**. Compositional trends of (a) $T_g$(x) (∇) and CS mode frequency ν(x) (●), (b) Molar volumes results from present work (●) , from ref [5] (■) and from ref. [4](Δ), (c) Non-reversing enthalpy at $T_g$ , $\Delta H_{nr}$(x), in present samples in fresh (**F**) state (∇) , after 2 weeks of aging at 25°C (○) curve **A1**, after 2 weeks of aging at 240°C (□) curve A2, and results from ref [7] (◊) curve **A0** after 2 weeks of aging at 25°C.The shaded panel gives the Intermediate Phase. The ▼ data points in the three panels correspond to wet samples. See text.

**Fig.4**. Elastic threshold compositions ($x_c$) and optical elastic power-laws (p) in (a) stressed-rigid and (b) Intermediate Phase deduced from the fitting the Raman mode frequency, $\nu_{CS}$(x), to equation 1.

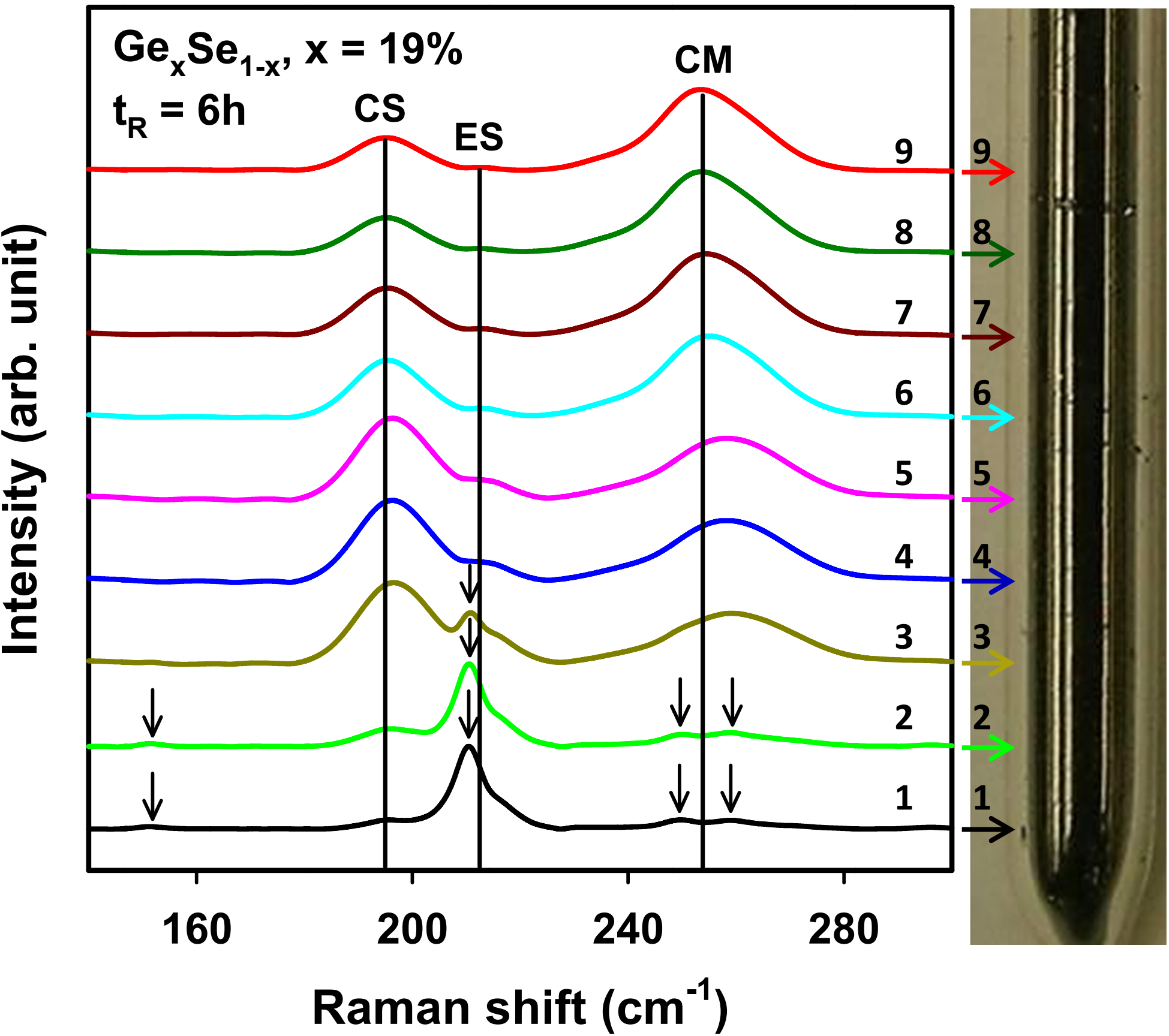

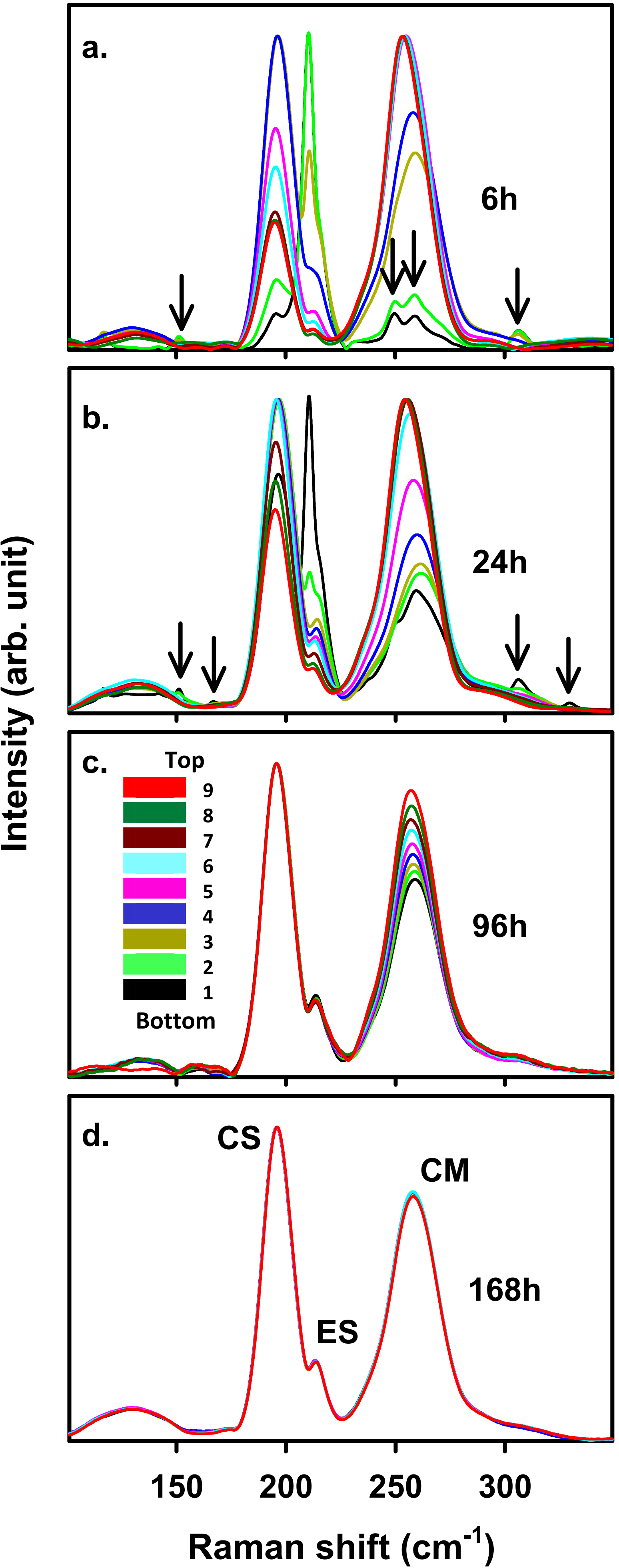

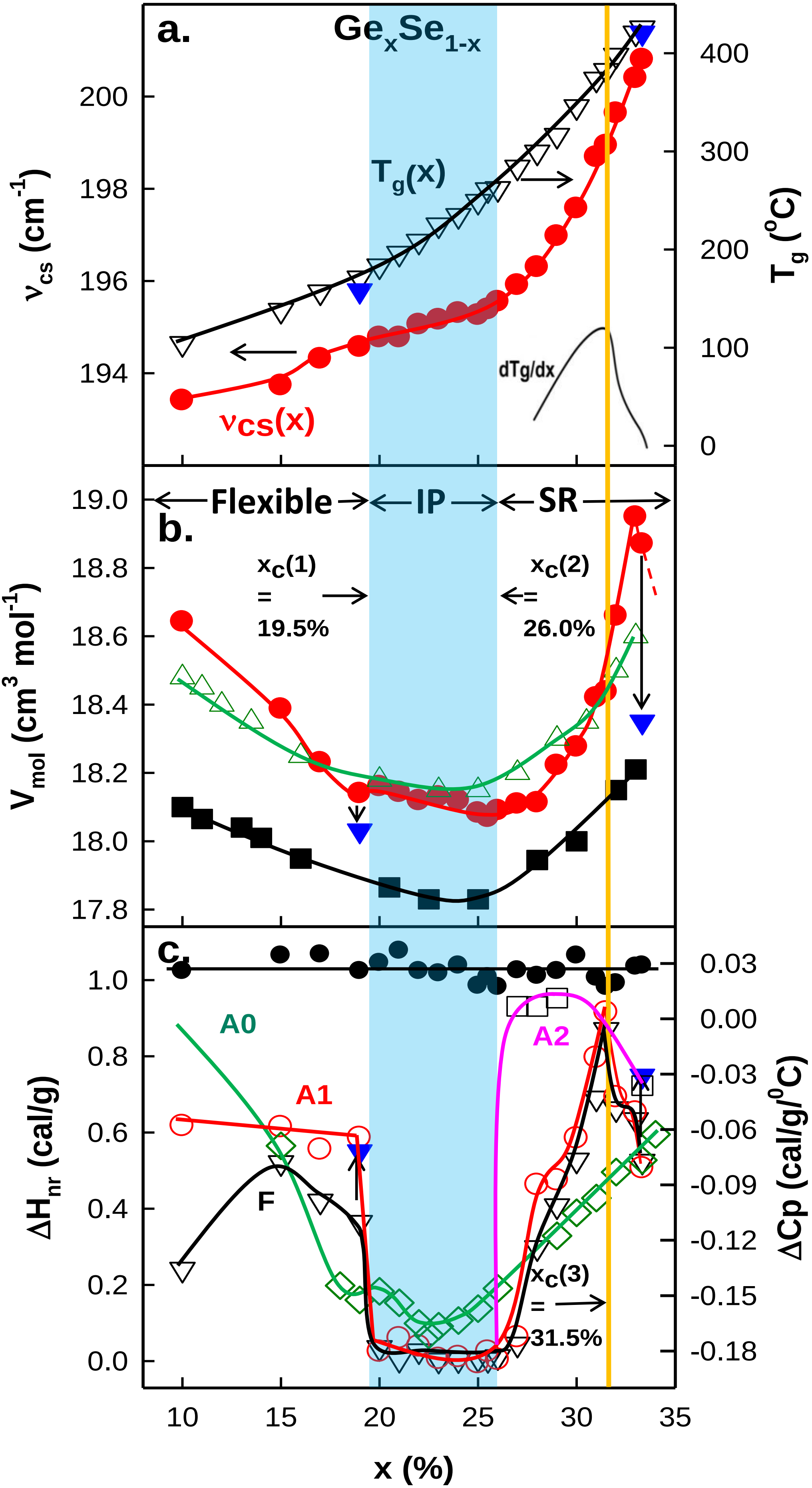

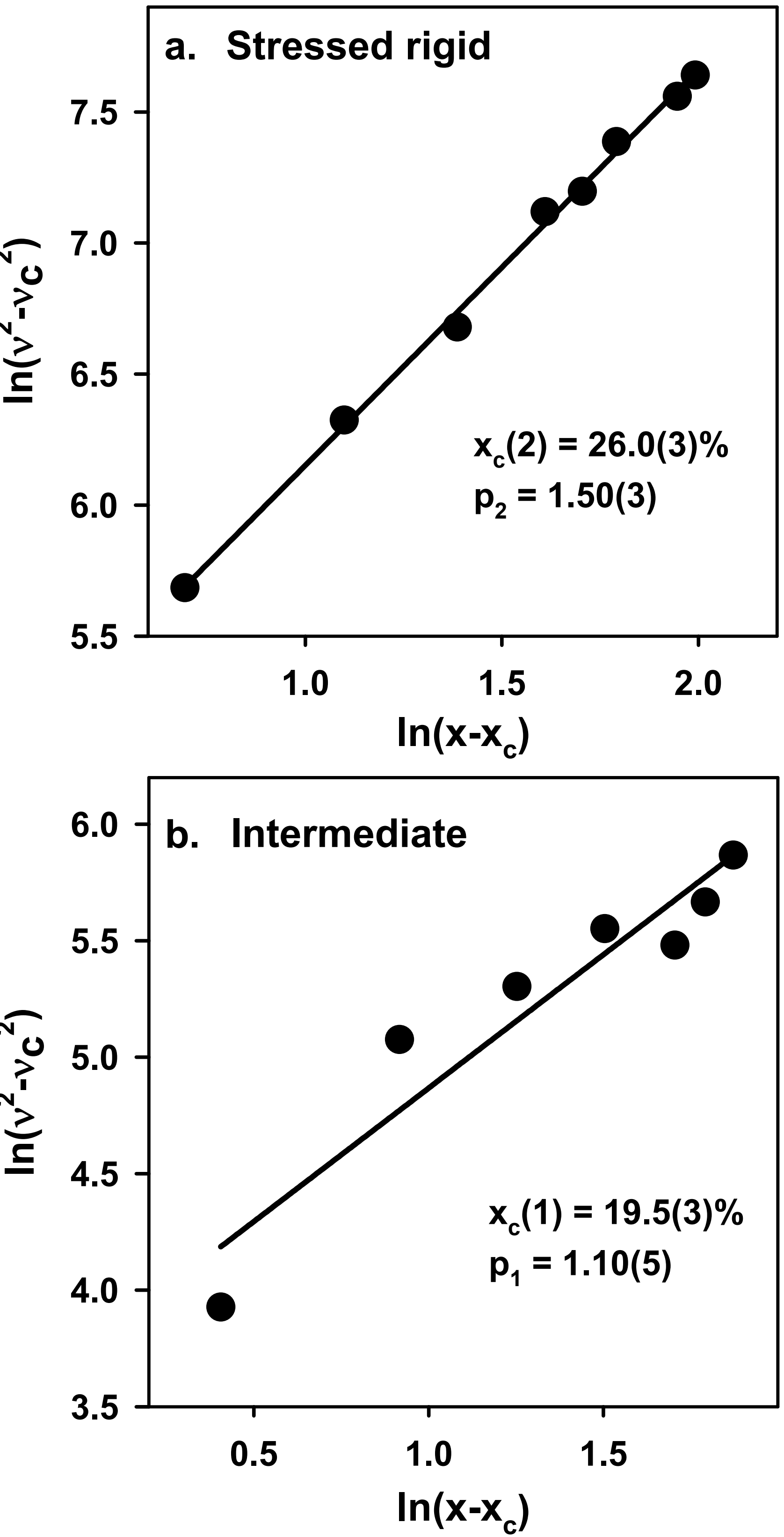